\documentclass[twocolumn,3p,preprint]{elsarticle}
\journal{Physics Letters B}
\usepackage[numbers]{natbib}
\RequirePackage{fancyhdr}
\RequirePackage{graphicx}
\usepackage{placeins}
\RequirePackage{adjustbox}
\usepackage{rotating}
\usepackage{multirow}
\usepackage{geometry}
\usepackage{ragged2e}
\usepackage{color}
\usepackage{array}
\usepackage{float}
\usepackage{units} 
\RequirePackage{amssymb}
\RequirePackage{epstopdf}
\RequirePackage{amsmath, amsfonts}
\biboptions{sort&compress}
\bibpunct{[}{]}{,}{n}{,}{,}

\begin{document}
\begin{frontmatter}

\title{Evolution of the dipole polarizability in the stable tin isotope chain}

\author[TUDarm]{S.~Bassauer}\ead{sbassauer@ikp.tu-darmstadt.de}
\author[TUDarm]{P.~von~Neumann-Cosel\corref{cor1}}\ead{vnc@ikp.tu-darmstadt.de}
\author[Erlangen]{P.-G.~Reinhard}
\author[RCNP]{A.~Tamii}
\author[RCNP]{S.~Adachi}
\author[TAM]{C.A.~Bertulani}
\author[RCNP]{P.Y.~Chan}
\author[Milano]{G.~Col\`o}
\author[TUDarm]{A.~D'Alessio}
\author[IOT]{H.~Fujioka}
\author[RCNP]{H.~Fujita}
\author[RCNP]{Y.~Fujita}
\author[RCNP]{G.~Gey}
\author[TUDarm]{M.~Hilcker}
\author[RCNP]{T.H.~Hoang}
\author[RCNP]{A.~Inoue}
\author[TUDarm,RCNP]{J.~Isaak}
\author[RIKEN]{C.~Iwamoto}
\author[TUDarm]{T.~Klaus}
\author[RCNP]{N.~Kobayashi}
\author[Miyazaki]{Y.~Maeda}
\author[Tohoku]{M.~Matsuda}
\author[TUDarm]{N.~Nakatsuka}
\author[NSCL]{S.~Noji}
\author[IMP,RCNP]{H.J.~Ong}
\author[Okayama]{I.~Ou}
\author[Zagreb]{N.~Paar}
\author[TUDarm]{N.~Pietralla}
\author[TUDarm]{V.Yu.~Ponomarev}
\author[Akal]{M.S.~Reen}
\author[TUDarm]{A.~Richter}
\author[Milano]{X.~Roca-Maza}
\author[TUDarm]{M.~Singer}
\author[TUDarm]{G.~Steinhilber}
\author[RCNP]{T.~Sudo}
\author[Rikkyo]{Y.~Togano}
\author[Kyoto]{M.~Tsumura}
\author[Tokyo]{Y.~Watanabe}
\author[TUDarm]{V.~Werner}

\address[TUDarm]{Institut f\"ur Kernphysik, Technische Universit\"at Darmstadt, D-64289 Darmstadt, Germany}
\address[Erlangen]{Institut f\"ur Theoretische Physik II, Universit\"at Erlangen, D-91058 Erlangen, Germany}
\address[RCNP]{Research Center for Nuclear Physics, Osaka University, Ibaraki, Osaka 567-0047, Japan}
\address[TAM]{Department of Physics and Astronomy, Texas A\&M University-Commerce, Commerce, Texas 75429, USA}
\address[Milano]{Dipartimento di Fisica, Universit\`a degli Studi di Milano and INFN, Sezione di Milano, 20133 Milano, Italy}
\address[IOT]{Department of Physics, Tokyo Institute of Technology, Tokyo 152-8551, Japan}
\address[RIKEN]{RIKEN, Nishina Center for Accelerator-Based Science, 2-1 Hirosawa, 351-0198 Wako, Saitama, Japan}
\address[Miyazaki]{Department of Applied Physics, Miyazaki University, Miyazaki 889-2192, Japan}
\address[Tohoku]{Department of Communications Engineering, Graduate School of Engineering, Tohoku University, Aramaki Aza Aoba, Aoba-ku, Sendai 980-8579, Japan}
\address[NSCL]{National Superconducting Cyclotron Laboratory, Michigan State University, East Lansing, Michigan 48824, USA}
\address[IMP]{Institute of Modern Physics, Chinese Academy of Sciences, Lanzhou, 730000, China}
\address[Okayama]{Okayama University, Okayama 700-8530, Japan}
\address[Zagreb]{Department of Physics, Faculty of Science, University of Zagreb, Zagreb, Croatia}
\address[Akal]{Department of Physics, Akal University, Talwandi Sabo, Bathinda Punjab-151 302, India}
\address[Rikkyo]{Department of Physics, Rikkyo University, Tokyo  171-8501, Japan}
\address[Kyoto]{Department of Physics, Kyoto University, Kyoto 606-8502, Japan}
\address[Tokyo]{Department of Physics, University of Tokyo, Tokyo 113-8654, Japan}


\cortext[cor1]{Corresponding author}

\begin{abstract}

The dipole polarizability of stable even-mass tin isotopes $^{112,114,116,118,120,124}$Sn was extracted from inelastic proton scattering experiments at \unit[295]{MeV} under very forward angles performed at RCNP.
Predictions from energy density functionals cannot account for the present data and the polarizability of $^{208}$Pb simultaneously.
The evolution of the polarizabilities in neighboring isotopes indicates a kink at $^{120}$Sn while all model results show a nearly linear increase with mass number  after inclusion of pairing corrections.
 
\end{abstract}

\begin{keyword}
$^{112,114,116,118,120,124}$Sn(p,p$^\prime$),
$E_{\rm p} = \unit[295]{MeV}$, 
$\theta_{\rm lab} = 0^\circ - 6^\circ$, 
relativistic Coulomb excitation, 
photoabsorption cross sections, 
dipole polarizability
\end{keyword}

\end{frontmatter}

\section{Introduction}
\label{sec:intro}

Determination of the nuclear Equation of State (EoS) is one of the major goals of current nuclear physics research \cite{roc18}, both experimentally and theoretically.  
Its knowledge is e.g.\ required for an understanding of astrophysical events like core-collapse supernovae \cite{yas20} or the formation \cite{oze16} or the mass and radius \cite{bog19a,bog19b} of neutron stars.  
In particular, the observation of a neutron star merger through the detection of gravitational waves \cite{abb17} and the associated electromagnetic spectrum provides a multitude of new experimental information, whose interpretation crucially depends on the EoS of neutron-rich matter \cite{fat18,cha19}.

The largest uncertainty of the EoS of proton-neutron asymmetric matter stems from the symmetry energy term.
Since the symmetry energy cannot be measured directly, experimental observables are sought that show a close correlation with its properties.  
The two most promising identified so far are the thickness of the neutron skin formed in heavy nuclei and the dipole polarizability, see e.g.\ Ref.~\cite{thi19}.  
The volume terms of nuclear energy density functional (EDF) theory -- presently the most successful approach to the microscopic description of heavy nuclei -- are directly related to nuclear bulk parameters such as the incompressibility $K$ or the symmetry energy $J$ and those bulk parameters often have a near one-to-one correspondence to nuclear observables.  
In particular, the symmetry energy $J$ shows a strong correlation with the slope of symmetry energy $L$, the neutron skin thickness ($r_n-r_p$), and the dipole polarizability $\alpha_D$ which has attracted much attention \cite{roc18,rei10,Klu09a,Pie12b,roc13,Naz14d,has15,Erl15b,mon16}.  
Accordingly, there is renewed interest in the measurement of the electric dipole strength or the corresponding photoabsorption cross sections in nuclei for an extraction of the dipole polarizability $\alpha_{\rm D}$ from inverse moments of the E1 sum rule \cite{boh81}
\begin{equation}
\label{eq:pol}
  \alpha_\mathrm{D}
  =
  \frac{\hbar c}{2\pi^{2} } 
  \int \frac{\sigma_\mathrm{abs}}{E_{\rm x}^{2}}{\rm d}E_{\rm x} 
  = 
   \frac{8 \pi}{9} \int\frac{\rm{B(E1)}}{E_{\rm x}}{\rm d}E_{\rm x},
\end{equation}
where $E_{\rm x}$ is the excitation energy, B(E1) the reduced electric dipole transition strength and $\sigma_{\rm abs}$ the photoabsorption cross section.  

In principle, the determination of $\alpha_{\rm D}$ requires data at all excitation energies.
However, It is well known from extensive studies in the past \cite{ber75,die88} that most of the E1 strength is concentrated in the IsoVector Giant Dipole Resonance (IVGDR).
Furthermore, the contribution from the high-energy region above the IVGDR is diminished by the inverse energy weighting in Eq.~(\ref{eq:pol}).
On the other hand, the role of low-energy strength is enhanced.
Heavy nuclei show resonance-like structures of isovector E1 strength below the IVGDR, typically around the neutron threshold, often called Pygmy Dipole Resonance (PDR). 
The PDR is observed in nuclei with neutron excess and thought to originate from those outermost neutrons that  display a soft spatial correlation with respect to the other nucleons forming the core of the nucleus under study.
This feature points towards a sensitivity of the Energy-Weighted Sum Rule (EWSR) exhausted by the PDR on the neutron pressure below saturation density and thus to a correlation with the properties of the EoS. 
However, the structure underlying the PDR and the resulting properties are not systematically understood yet \cite{sav13,bra19} and a simple relation to bulk properties is questionable~\cite{Rei13b}.

Recently, inelastic proton scattering at energies of a few hundred MeV and very forward angles including $0^\circ$ has been established as a new method to extract the complete E1 strength in heavy nuclei from low excitation energies across the giant resonance region \cite{vnc19}.  
Under this particular kinematics selective excitation of E1 and spin-M1 dipole modes is observed.  
Their contributions to the cross sections can be separated either by a Multipole Decomposition Analysis (MDA) of the cross sections  \cite{pol12} or independently by the measurement of a combination of polarization transfer observables \cite{vnc19,tam11}.  
Good agreement of both methods was demonstrated for reference cases \cite{tam11,has15,mar17} indicating that the much simpler measurement of cross sections using an unpolarized beam and employing the MDA thereof is sufficient for a reliable extraction of the E1 strength distribution.

Two different driving agents for the evolution of the dipole polarizability are conceivable, viz.\ neutron excess and the general trend with mass number $A$ (i.e., the size) both dependent on the symmetry energy.
The chain of proton-magic tin nuclei is of particular interest because the underlying structure changes little between neutron shell closures $N = 50$ and 82 and thus highlights the role of neutron excess.  
Accordingly, a variety of model calculations have been performed attempting to explore this connection
\cite{tso04,vre04,pie06,kam06,ter06,tso08,lit08,lan09,lit10,dao11,avd11,Pie12b,pap14,pie14,Naz14d,eba14,Erl15b,yuk19}.
While the correlation of $\alpha_{\rm D}$, $J$ and $L$ in EDF models is robust \cite{roc13}, quantitative predictions differ considerably because the static isovector properties of the interactions are usually poorly constrained by the data used to determine the interaction parameters.  
The present letter reports on a systematic study of the dipole polarizability in the stable even-mass tin isotopes and provides an important benchmark for the attempts to develop interactions with predictive power as a function of nuclear mass and neutron excess.

\section{Experiment}
\label{sec:exp-results}

The experiments were performed at the Research Center for Nuclear Physics (RCNP), Osaka University, using the Grand Raiden Spectrometer \cite{fuj99}. 
The proton beam had an energy $E_{\rm p} = 295$ MeV.
Typical beam currents were between 2 and \unit[20]{nA}, depending on the spectrometer angle. 
Data were taken at central spectrometer angles $0^\circ$, $2.5^\circ$ and $4.5^\circ$.
Highly enriched self-supporting targets of $^{112,114,116,118,120,124}$Sn with areal densities between 3 and \unit[7]{mg/cm$^2$} were used.
The dispersion-matching technique enabled measurements with energy resolutions between 30 and 40 keV (full width at half maximum, FWHM).
The experimental techniques and the raw data analysis are described in Ref.~\cite{tam09} and specific to the present experiment in Ref.~\cite{bas20}.
Data taking for $^{120}$Sn was limited to a cross check of experimental cross sections obtained in a previous experiment \cite{has15,kru15}. 

\begin{figure}[t]
\begin{center}
	\includegraphics[width=\columnwidth]{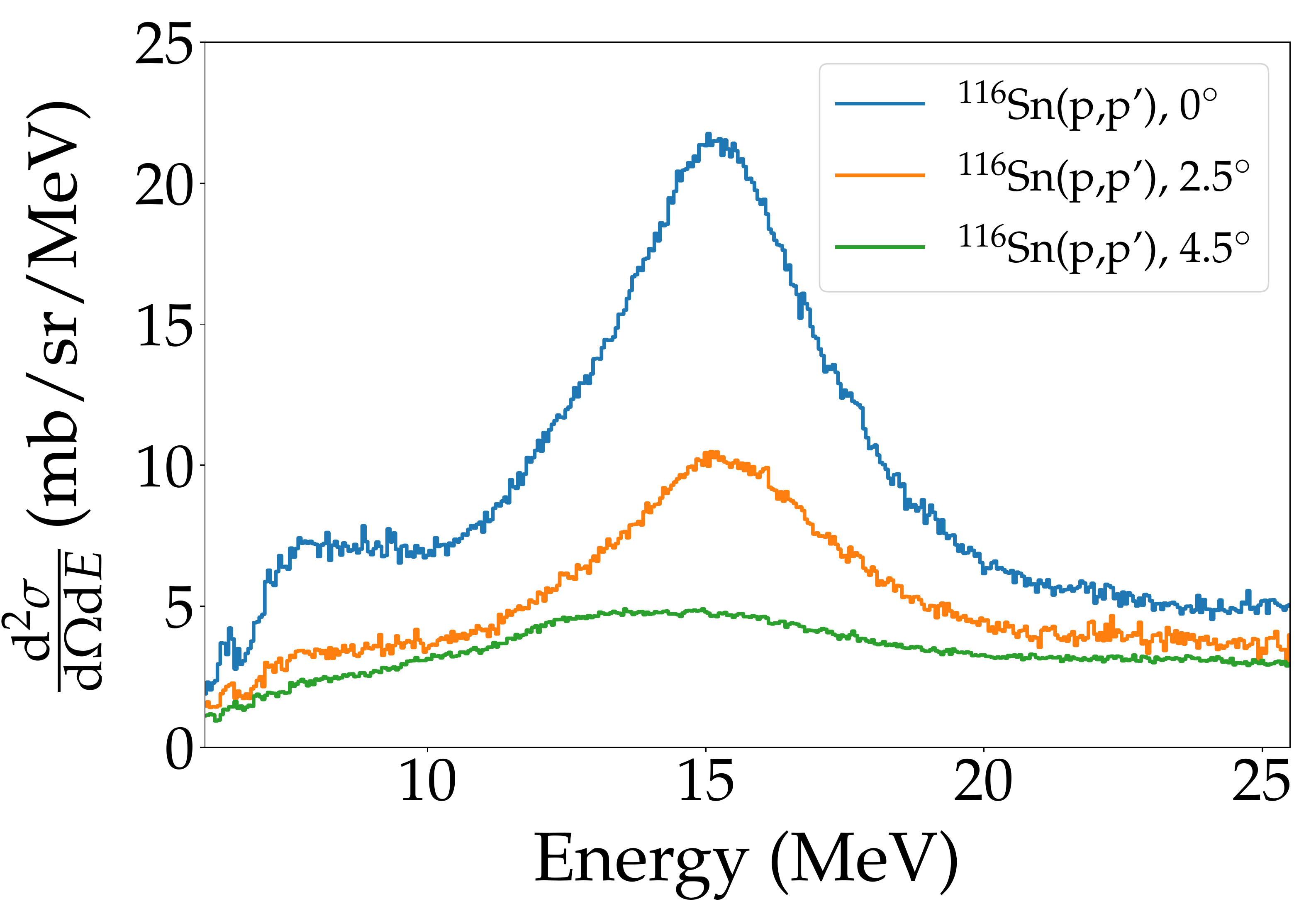}
	\caption{Double differential cross section of the $^{116}$Sn(p,p$^\prime$) reaction at \unit[$E_{\rm p} = 295$]{MeV} and scattering angles $\Theta_{\rm lab} = 0^\circ$, $2.5^\circ$ and $4.5^\circ$. 
	}
	\label{fig:Spectrum}
\end{center}
\end{figure}
Typical spectra at the three main spectrometer angles are shown in Fig.~\ref{fig:Spectrum} for $^{116}$Sn by way of example.
The dominance of relativistic Coulomb excitation expected for the kinematics at scattering angles close to $0^\circ$  \cite{vnc19} suggest that the prominent excitation centered at about 15 MeV is due to the IVGDR.
At lower excitation energies a pronounced structure is visible which also slowly disappears with increasing scattering angle.
The angular dependence indicates a dipole character of the excited states underlying this structure as demonstrated in the next section. 

\section{Multipole Decomposition Analysis}
\label{sec:MDA}

An MDA of the cross section angular distributions was performed based on a least-squares fit of the type 
\begin{equation}
 \sum_{i} 
 \left(\frac{\textnormal{d}\sigma}{\textnormal{d}\Omega}(\Theta_i,E_{\rm x})_{\textnormal{exp}} - \frac{\textnormal{d}\sigma}{\textnormal{d}\Omega}(\Theta_i,E_{\rm x})_{\textnormal{th}}\right)^2 \equiv {\rm min},
\label{eq:fit}
\end{equation}
where
\begin{align}
\frac{\textnormal{d}\sigma}{\textnormal{d}\Omega}(\Theta,E_{\rm x})_{\textnormal{th}} =&
\sum_{O \lambda} a_{O \lambda}  \frac{\textnormal{d}\sigma}{\textnormal{d}\Omega}(\Theta,E_{\rm x},O\lambda)_{\textnormal{DWBA}} \nonumber \\
& +b\frac{\textnormal{d}\sigma}{\textnormal{d}\Omega}(\Theta)_{\textnormal{QFS}},
\label{eq:theo}
\end{align}
with the condition that all coefficients $a_{O \lambda}$ and $b$ were positive.
The spectra were analyzed in \unit[200]{keV} bins.
The angular acceptance of the spectrometer of $\pm 2.7^\circ$ allowed to generate five data points per angle and energy bin, so that in total 15 data points between $0.8^\circ$ and $5.5^\circ$ were available for the MDA.
The shapes of theoretical angular distributions for different electric ($E)$ and magnetic ($M$) multipolarities $O \lambda$ ($O = E,M)$ calculated with the code DWBA07 \cite{dwba07} and based on transition densities from the Quasiparticle Phonon Model (QPM) \cite{sol92}  were used as input. 
As demonstrated for previous cases \cite{pol12, mar17}, the low momentum transfers of the experiment permit a restriction of multipoles in Eq.~(\ref{eq:theo}) to E1, M1 and one multipole representing all contributions $\lambda > 1$ (E3 in the present case).
Above the particle threshold the spectra contain a phenomenological background dominated by quasifree scattering (QFS).
Its angular distribution shape determined at the highest excitation energies measured ($23-25$ MeV), where the IVGDR contributions are negligible, was found to be constant and assumed to also hold in the giant resonance region.

Prior to the MDA, the contributions of the isoscalar giant monopole and quadrupole resonances were subtracted from the spectra following the method described in Ref.~\cite{don18}, again using QPM results of the corresponding strengths.
The experimental strength distributions were taken from Ref.~\cite{li10}.

\begin{figure}[t]
\begin{center}
	\includegraphics[width=\columnwidth]{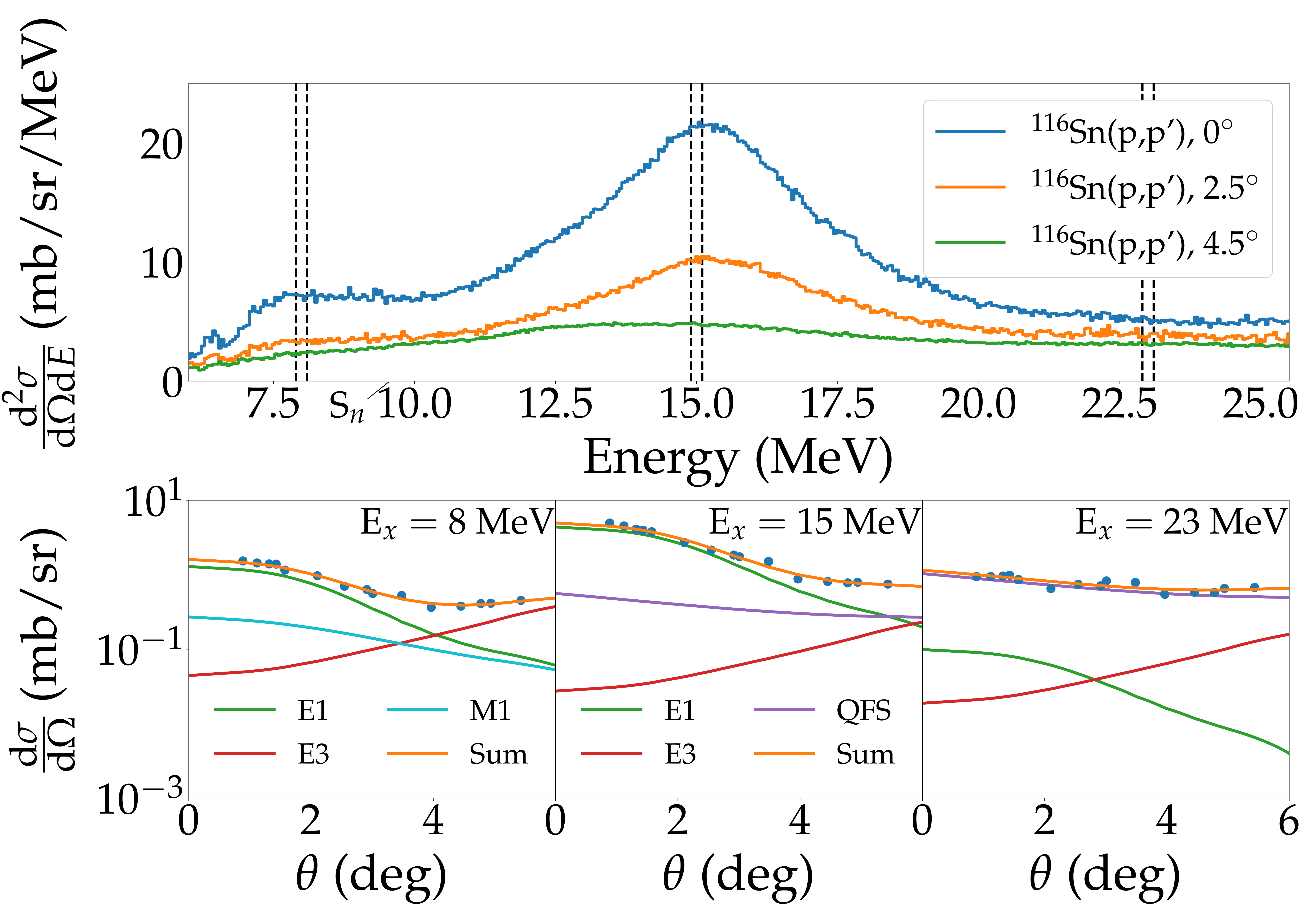}
	\caption{Results of the MDA for three different excitation energy bins at 8, 15 and \unit[23]{MeV} indicated by vertical dashed lines, respectively, shown for the example of $^{116}$Sn. 
	}
	\label{fig:MDA}
\end{center}
\end{figure}
Examples of the fits of Eq. (\ref{eq:fit}) are displayed in Fig.~\ref{fig:MDA} for the case of the $^{116}$Sn(p,p$^{\prime}$) data.
The upper part shows selected bins at excitation energies of 8, 15 and 23 MeV  indicated by the vertical dashed lines in the spectra, whose angular distributions are plotted in the lower part.
At the lowest excitation energy one finds a dominance of E1 cross sections close to $0^\circ$, but M1 contributions are non-negligible.
Above $4^\circ$ the spectra are dominated by $\lambda > 1$ multipoles. 
At 15 MeV near the maximum of the IVGDR one observes the expected dominance of E1 cross sections.
Finally, at 23 MeV the phenomenological background is most important and all other contributions are small.

\section{Photoabsorption cross sections}
\label{sec:gamcomp}

The Coulomb excitation cross sections resulting from the MDA were converted to equivalent photoabsorption cross sections using the virtual photon method~\cite{ber88}.  
The virtual photon spectrum was calculated in an eikonal approach~\cite{ber93} in contrast to the previous study of $^{120}$Sn, where the semiclassical approximation was used \cite{has15,kru15}.
In heavy nuclei the differences between both approaches are small (typically less than 10\%) but in lighter nuclei the semiclassical approach fails \cite{bir17}.   
A comparison of both approaches and the typical energy dependence for the present kinematic conditions are shown in Sec.~3.3 of Ref.~\cite{vnc19} for the example of $^{120}$Sn.
Although the experimental spectra extend above 20 MeV, the E1 cross sections become too small with respect to the quasifree background for a meaningful decomposition in the MDA.
\begin{figure}
\begin{center}
	\includegraphics[width=\columnwidth]{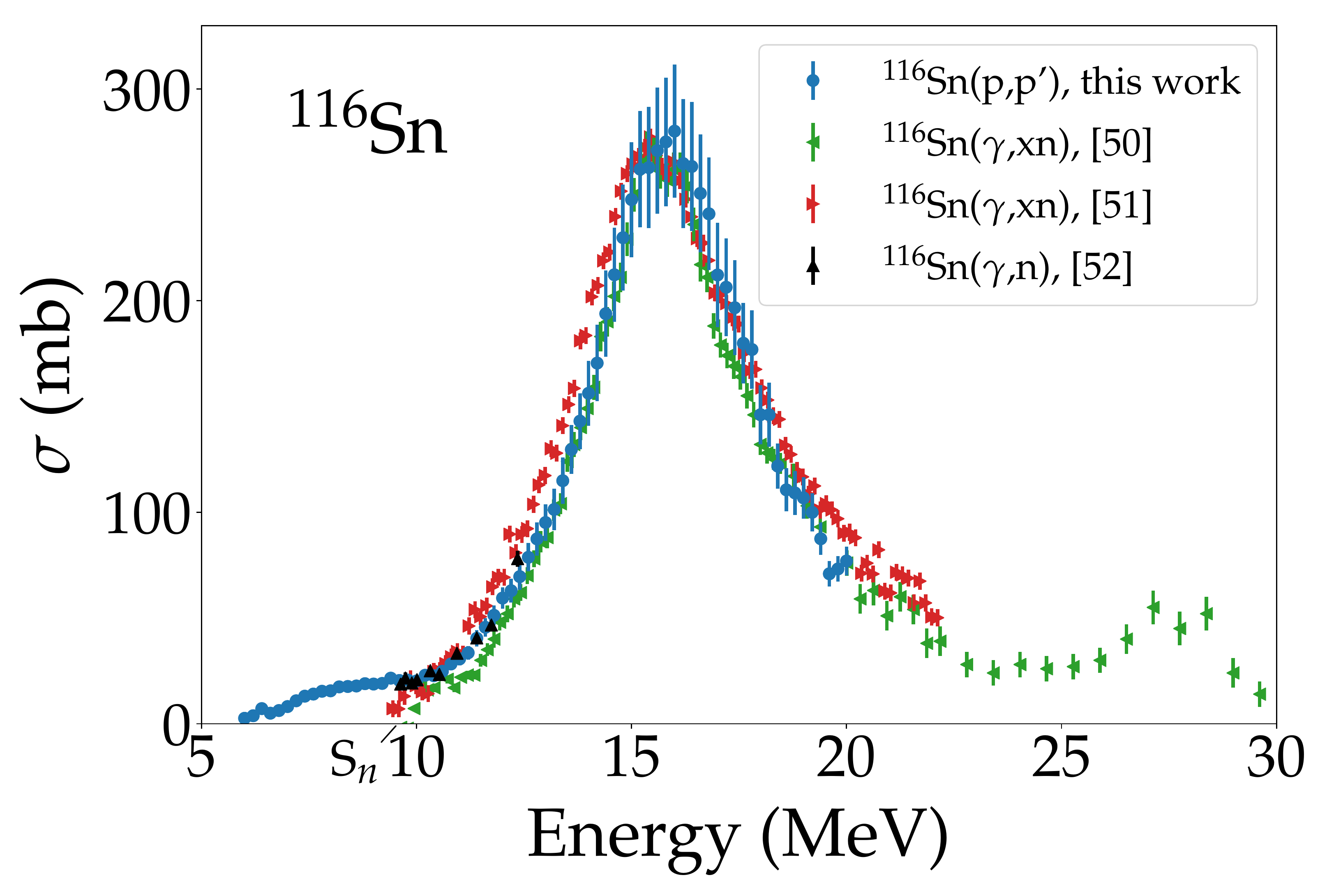}
	\caption{Photoabsorption cross sections of $^{116}$Sn deduced from the present experiment in comparison to previous work \cite{ful69,lep74,uts09}. 
	}
	\label{fig:PhotoCS}
\end{center}
\end{figure}
Figure \ref{fig:PhotoCS} presents a comparison of the resulting photoabsorption cross sections with data available from ($\gamma$,xn) experiments \cite{ful69,lep74,uts09}, again for the example of $^{116}$Sn.
The error bars include contributions from systematic uncertainties of the cross sections determination and from the MDA decomposition \cite{bas20} added quadratically.
Statistical errors are negligible.

One finds significant differences on the low-energy flank  of the IVGDR. 
The Livermore data by Fultz et al.~\cite{ful69} show the best agreement with the present result above 12 MeV, but undershoot the data from all other experiments for $E_{\rm x} < 12$ MeV.
Near neutron threshold the new ($\gamma$,n) data of Ref.~\cite{uts09} agree best with our results.
Similar differences are found for the other isotopes studied here.
A detailed account is given elsewhere \cite{bas20}.

\section{Dipole polarizability}
\label{sec:DP}

The present data provide photoabsorption cross sections in the energy region $6-20$ MeV for the determination of $\alpha_{\rm D}$ from Eq.~(\ref{eq:pol}).  
Below 6 MeV, B(E1) strength distributions are available for $^{112,116,120,124}$Sn from nuclear resonance fluorescence experiments \cite{oze14,gov98}, but were neglected for consistency with the other isotopes.  
These contributions are generally small (\unit[$<0.5$]{\%} of the total dipole polarizability).  
In Ref.~\cite{roc15} it was argued that the contributions of the quasideuteron mechanism \cite{sch88}, which dominates the photoabsorption for excitation energies above 30 MeV in the present case, should be excluded from the integration of Eq.~(\ref{eq:pol}).  
Such a nonresonant process is not included in the model calculations.  
Data are available from Ref.~\cite{ful69} in the excitation region $20 -30$ MeV for $^{116,118,120,124}$Sn.  However, we refrain from using them, since these results show large variations between different isotopes but no systematic isotopic dependence as discussed in Ref.~\cite{bas20}.
Including these data would not only alter the absolute values but also significantly modify the isotopic dependence.  
Rather we employ a theory-assisted estimate of strength in the region above 20 MeV.  
To that end, we performed calculations at the level of quasiparticle random phase approximation (QRPA) \cite{Rin80B} and of the more detailed QPM \cite{vnc19,rye02,pol12,tam11,pol14}. 
The QRPA and QPM cross sections used to calculate the dipole polarizability in the energy region above 20 MeV were convoluted with Lorentzians whose widths were tuned to reproduce the present IVGDR data.  
We have done that for different models and parametrizations and find consistently the same contribution of about \unit[8]{\%} to $\alpha_{\rm D}$.  
A particularly encouraging result is the most elaborate test based on a fully self-consistent continuum RPA calculation \cite{Tse17a} with the Skyrme parametrization SV-bas \cite{Klu09a}, for technical reasons performed for the doubly magic nucleus $^{132}$Sn. 
It provides a proper description of the experimental photoabsorption cross sections \cite{adr05} without further folding and finds a contribution of \unit[6]{\%} in the somewhat larger $^{132}$Sn.  
To account for the uncertainties in that extrapolation, an error of \unit[10]{\%} is associated with the contributions taken from the model results.

\begin{figure}
\begin{center}
	\includegraphics[width=\columnwidth]{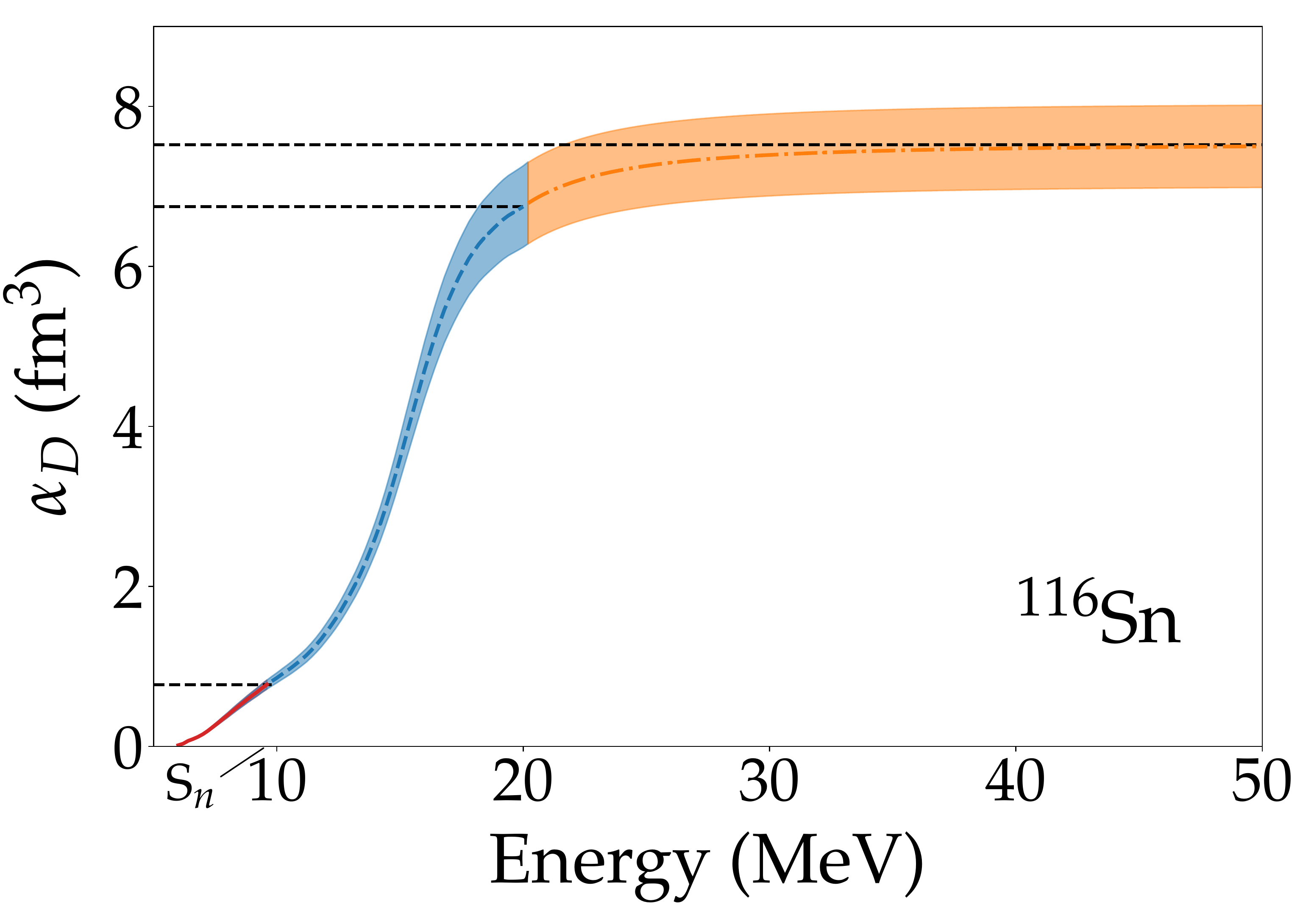}
	\caption{Running sum of the dipole polarizability from the present (p,p$^\prime$) data for the example of $^{116}$Sn.
	Red: Contribution from 6 MeV to $S_{\rm n}$. 
	Blue: Contribution from $S_{\rm n}$ to 20 MeV.
	Orange: Contribution above 20 MeV from QPM calculations, see text for details.
	}
	\label{fig:DPRunSum}
\end{center}
\end{figure}
Figure \ref{fig:DPRunSum} displays the evolution of $\alpha_{\rm D}$ as a function of excitation energy (the running sum) for the example of $^{116}$Sn.  
The error band considers statistical and systematical uncertainties, the latter including contributions from experiment and from the MDA (for details see Ref.~\cite{bas20}).
The figure demonstrates that the polarizability values are dominated by the contribution of the IVGDR (blue), but the low-energy (red) and high-energy (orange) parts are non-negligible.
The corresponding partial and total values are summarized in Table \ref{tab:pol}.  
The low-energy contribution up to the neutron threshold ($S_{\rm n}$) -- i.e.\ the part missed in ($\gamma$,xn) experiments -- varies from \unit[13]{\%} ($^{112}$Sn) to \unit[8]{\%} ($^{124}$Sn) due to the decrease of $S_{\rm n}$ as a function of mass number.  
The high-energy contribution from the QPM calculations amounts to \unit[$9-10$]{\%} in all isotopes.
\begin{table}
	\centering
	\caption{Total dipole polarizability $\alpha_{\rm D}$ of $^{112,114,116,118,120,124}$Sn (in fm$^3$) determined as described in the text.
	Partial values are given for the contributions from \unit[6]{MeV} to the neutron threshold $S_{\rm n}$, from $S_{\rm n}$ to \unit[20]{MeV} and \unit[$> 20$]{MeV}.
	}
	\begin{tabular}{ccccc}
 		\hline
 		\noalign{\vskip 0.5mm}
 		& $6-S_{\rm n}$ & $S_{\rm n}-20$ & $> 20$ & Total  \\
 		\hline
 		\noalign{\vskip 0.5mm}
 		$^{112}$Sn & $0.94(7)$ & $5.51(42)$ & $0.73(7)$ & $7.19(50)$ \\
 		$^{114}$Sn & $0.83(7)$ & $5.74(51)$ & $0.72(7)$ & $7.29(58)$ \\
 		$^{116}$Sn & $0.77(6)$ & $5.98(45)$ & $0.77(8)$ & $7.52(51)$ \\
 		$^{118}$Sn & $0.78(9)$ & $6.36(78)$ & $0.77(8)$ & $7.91(87)$ \\
 		$^{120}$Sn & $0.84(7)$ & $6.49(52)$ & $0.75(8)$ & $8.08(60)$ \\
 		$^{124}$Sn & $0.65(5)$ & $6.49(51)$ & $0.85(8)$ & $7.99(56)$ \\
    	\hline
	\end{tabular}
	\label{tab:pol}
\end{table}

We note that a larger value for $^{120}$Sn was published in Ref.~\cite{has15} based on the same type of experiment, which after correction for the quasideuteron part amounted to  \unit[$\alpha_\mathrm{D} = 8.59(37)$]{fm$^3$}.
However, the difference to the present result is not due to the (p,p$^\prime$) data (cross sections from the previous and present experiments agree within error bars), but result from averaging in Ref.~\cite{has15} with the ($\gamma$,xn) data of Refs.~\cite{ful69,lep74}, whose contributions to $\alpha_{\rm D}$ from the IVGDR region are larger than from the present work, and from the particularly large photoabsorption strengths of Ref.~\cite{ful69} in the energy region $20-30$ MeV for the case of $^{120}$Sn (see Ref.~\cite{bas20}). 

The new polarizability results are now discussed in comparison to theoretical predictions from nuclear EDFs (for a general review see Ref.~\cite{Ben03aR}) based on the non-relativistic Skyrme functional and the relativistic mean field model (RMF). 
As argued in already in the introduction, dipole polarizability $\alpha_\mathrm{D}$ is a key observable for probing the symmetry energy which, in turn, is important to determine the nuclear EoS.
The present data on $\alpha_\mathrm{D}$ in the Sn isotopic chain provide new insights to that discussion, in particular on the decomposition of the role of neutron excess vs.\ the global mass dependence.  

We have scrutinized a great variety of published EDF parametrizations, but confine the present discussion to four typical representatives:
SV-bas is a Skyrme functional tuned to a large pool of ground state properties (energy, radius, surface thickness) with additional constraints on the EoS such that it reproduces giant resonances and $\alpha_\mathrm{D}$ in $^{208}$Pb \cite{Klu09a}. 
KDE0-J33 \cite{agr05} is also a Skyrme functional with fewer ground state properties in the fit pool and more bias on bulk properties; we take here the one sample from the KDE0 series closest to the present results.  
DDMEa is a RMF functional with density-dependent meson coupling \cite{vre03}. 
Finally, DD-PCX is a different fit to RMF \cite{yuk19} tuned to reproduce $\alpha_\mathrm{D}(^{208}\mathrm{Pb})$ \cite{tam11} after correction for the quasideuteron part \cite{roc15}.  
In all cases, pairing is included in the calculations. 
We have chosen these four cases to cover relativistic as well as non-relativistic models and within each group two EDFs which differ sufficiently concerning other properties. A more detailed evaluation covering larger sets of EDFs will be given in a subsequent publication.

\begin{figure}
\begin{center}
	\includegraphics[width=\columnwidth]{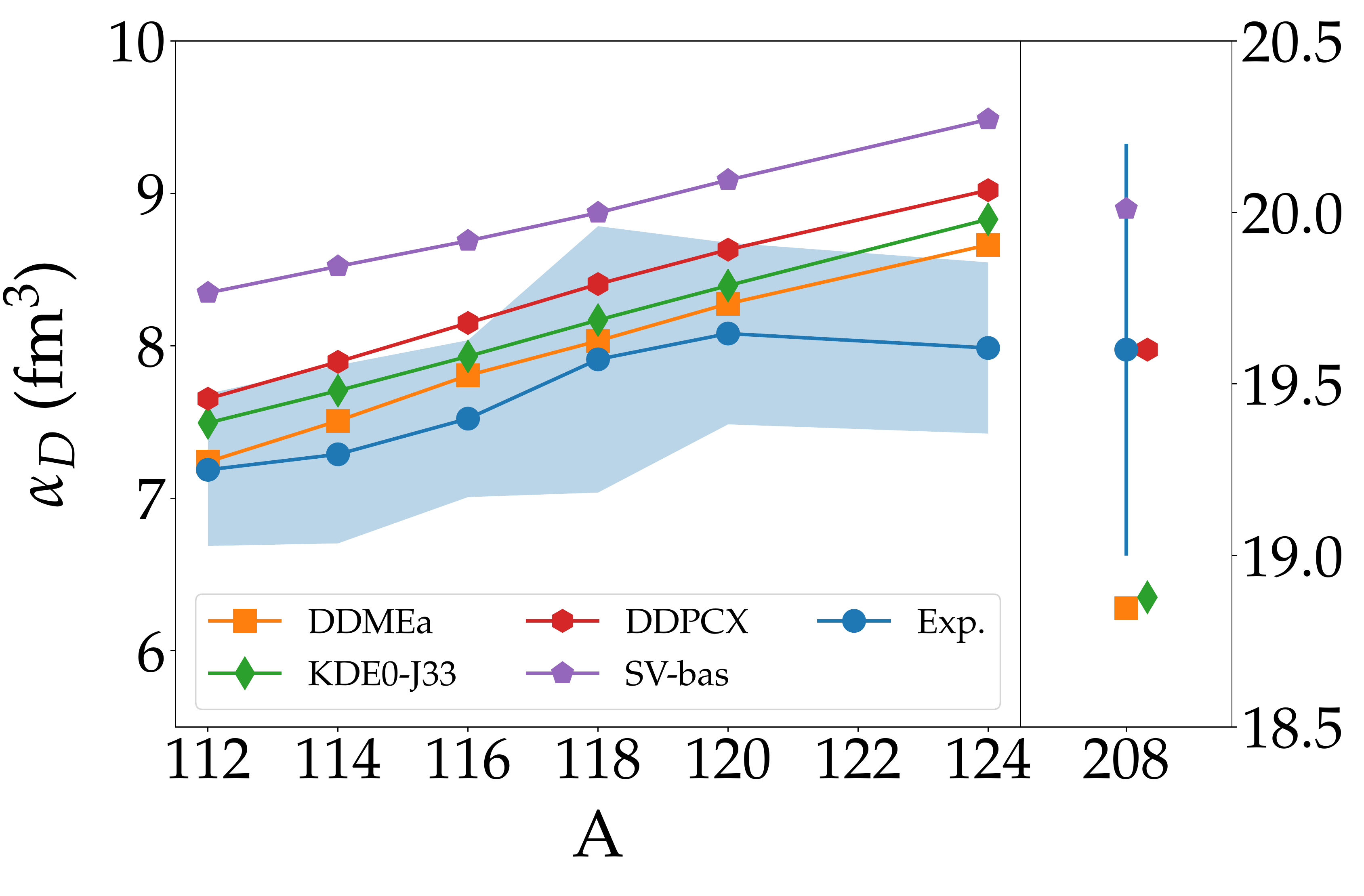}
	\caption{Dipole polarizability $\alpha_\mathrm {D}$ in the stable Sn isotopes (left panel) and in $^{208}$Pb (right panel). 
	Note the different scale for $^{208}$Pb.  
	The experimental values (blue dots) and their errors (blue band) are compared with theoretical results from the Skyrme functionals KDE0-J33 \cite{agr05,mon16} (green diamonds) and SV-bas \cite{Klu09a} (purple pentagons) and the RMF functionals DDMEa \cite{vre03} (orange squares) and DD-PCX \cite{yuk19} (red hexagons). 
	}
	\label{fig:AlphaExpTheoAll}
\end{center}
\end{figure}
Figure \ref{fig:AlphaExpTheoAll} shows $\alpha_\mathrm{D}$ in the Sn chain (left) and $^{208}$Pb (right) comparing data with results from the four selected EDF parametrizations.
At first glance, all four EDFs lie reasonably close to all data.  
The same holds for most of the other more recent, well tuned EDFs because isovector trends of ground state data imprint already some information on the isovector response.
A closer look reveals interesting differences from which we may learn about nuclear response properties.
SV-bas was tuned to the value of $\alpha_\mathrm{D}$ in $^{208}$Pb \cite{tam11} prior to the correction for the quasideuteron part and found to perform very well for the older, larger value of $\alpha_\mathrm {D}$ in $^{120}$Sn in Ref.~\cite{has15}, but now lies above the values of the present work.  
DD-PCX was tuned to $\alpha_\mathrm{D}(^{208}\mathrm{Pb})$ after correction for the quasideuteron part \cite{roc15} and thus fits perfectly for $^{208}\mathrm{Pb}$ while being somewhat high for Sn.  
KDE0-J33 and DDMEa  come closest for the Sn chain, however at the price of underestimating $\alpha_\mathrm{D}(^{208}\mathrm{Pb})$.

The similarity of the ordering along the Sn chain and in $^{208}$Pb shows that one can shift the $\alpha_\mathrm{D}$ values globally up and down without sacrificing too much of the overall quality of a functional, a feature observed already in earlier studies (see e.g.\ Refs.~\cite{Klu09a,Pie12b,mon16}).  
The trend with nucleon number $A$ looks rather rigid leading to very similar slopes along the Sn chain and, on a wider scale, to strict relations between $^{208}$Pb and the Sn isotopes.  
This is, in fact, expected from Migdal's hydrodynamical model \cite{mig45}. 
The rigidity of the trends with $A$ poses an intriguing problem for the given functionals: 
it seems that one cannot accommodate the $\alpha_\mathrm{D}$ data in $^{208}$Pb and in the Sn isotopes simultaneously. 
For a quantitative estimate we can look at statistical correlations between observables, see e.g.~Refs.~\cite{rei10,Dob14a}. 
The correlations of $\alpha_\mathrm{D}$ within the present Sn chain are better than 97\%. 
The correlation of of $\alpha_\mathrm{D}$ between $^{208}$Pb and the Sn isotopes is with still large (about 90\%), but already less restrictive.  
We can also see from Fig.~\ref{fig:AlphaExpTheoAll} that the $A$ dependence of $\alpha_\mathrm{D}$ allows some variations as the slope of the four EDFs is not fully identical.  
We have also checked the large scale trend from $^{208}$Pb to Sn with other EDFs and find that the density dependence of a functional plays a role, as already indicated in previous studies with both RMF \cite{vre03,pie11,Naz14d} and Skyrme functionals \cite{Erl10a}). 
This sets the direction for future improvement.  
We have to develop EDFs describing $\alpha_\mathrm{D}$ over a wide range of $A$ including also data on $^{40}$Ca \cite{fea20}, $^{48}$Ca \cite{bir17} and $^{68}$Ni \cite{ros13} and $^{90}$Zr \cite{kla20} into the fit, first trying to exploit the leeway in present functionals and, if necessary, to consider EDFs with more elaborated density dependencies.  
Such large scale fits should, moreover, reduce the extrapolation uncertainties for the symmetry energy $J$ which are about 1.4 MeV with the present error bands for
$\alpha_\mathrm{D}(^{208}\mathrm{Pb})$ \cite{roc15}. 
The description of the IVGDR, where a similar problem in describing the trend with $A$ exists \cite{Erl10b}, is also expected to benefit.

A further aspect of interest in Fig.~\ref{fig:AlphaExpTheoAll} are the trends with $A$ depicted in the left panel.  
The four theoretical results shows a similar nearly linear behavior while the experimental data slightly bend over around $^{120}$Sn. 
Recall that $^{120}$Sn corresponds to a subshell closure (below the neutron $1h_{11/2}$ shell).
Indeed, calculations with the RMF parametrization FSU040
\cite{tod05,pie11}, which uses the filling approximation rather than pairing, delivers qualitatively the experimental trend.  
This indicates that the pairing strength plays a role for details of the trend caused by shell effects.  
Thus, the role of pairing deserves further careful investigations.

\begin{figure}
\begin{center}
	\includegraphics[width=\columnwidth]{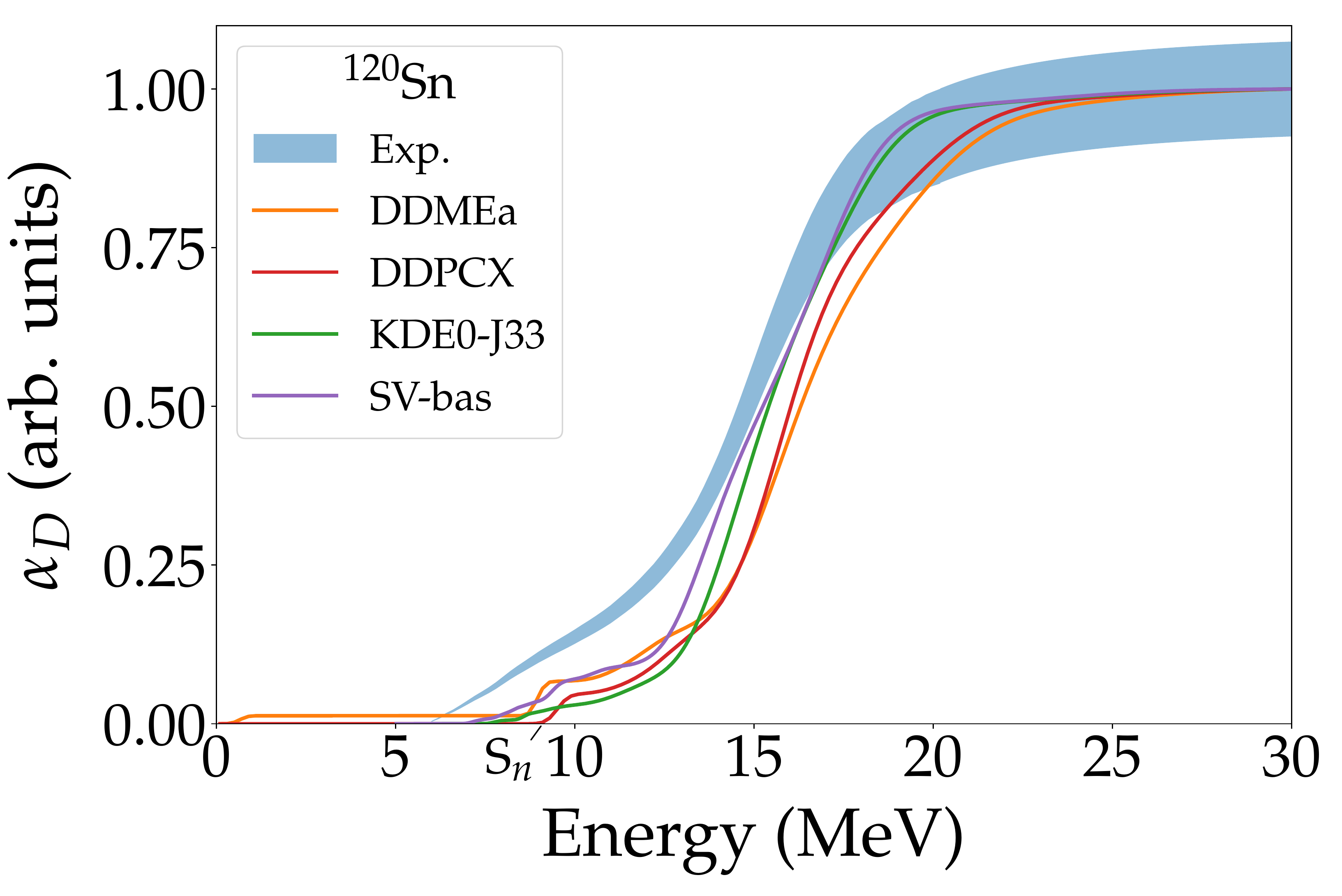}
	\caption{Normalized running sums of the dipole polarizability in $^{120}$Sn of the four EDFs under consideration in comparison to the experimental data. 
The theoretical results from discrete RPA were smoothed by Lorentzians to account for escape and spreading width.  
}
	\label{fig:120SnRunSum}
\end{center}
\end{figure}
We finally address possible relations between low-lying dipole strength in the PDR regime and $\alpha_\mathrm{D}$ discussed in the introduction.  
The inverse energy weights in the integration to $\alpha_\mathrm{D}$ in Eq.~(\ref{eq:pol}) emphasizes the low-energy region. 
In order to quantify its contribution, we show in Fig.~\ref{fig:120SnRunSum} a comparison of of the normalized experimental and theoretical running sums of $\alpha_\mathrm{D}$ for the example of $^{120}$Sn. 
The bounds of the PDR region in the calculations are not sharply defined, but should lie near 11 MeV. 
There are some variations of the predicted contribution but all are well below the experimental value of about 15\% in that energy region.
Note that the experimental curve is much softer at the lower and upper ends. 
The empirical smoothing of the RPA spectra simulating line broadening from many-body configurations is obviously insufficient in the wings of the spectra. 
This, in turn, indicates that a large part of dipole strength observed in the PDR region can be interpreted as a low-energy tail of the IVGDR \cite{kru15}. 
A clean separation of the two components requires many-body calculations beyond RPA.

\section{Conclusions}
\label{sec:conclusion}

We have extracted the dipole polarizability of stable even-mass Sn isotopes from relativistic Coulomb excitation using 295 MeV inelastic proton scattering at very forward angles. 
This allows to deduce precise data on the photoabsorption cross section up to 20 MeV. 
The technique provides, in particular, data indpendent of the neutron emission threshold. 
The results permit detailed studies of isotopic trends of isovector properties of nuclei carried in the IVGDR, the dipole polarizability $\alpha_\mathrm{D}$, and the 
low-lying dipole strength (PDR) \cite{bas20}.

We have exemplified the potential of the new data with a brief discussion of the case of the dipole polarizability $\alpha_\mathrm{D}$, an observable whose direct relation to isovector bulk properties (symmetry energy) makes it particularly important for theoretical developments.
Although practically all up-to-date EDF parametrizations provide at once roughly acceptable values for $\alpha_\mathrm{D}$, there are instructive differences in detail.  
The new $\alpha_\mathrm{D}$ values are systematically lower than the old value for $^{120}$Sn which calls for a new fine-tuning of EDF parametrizations. 
Furthermore, comparison with $\alpha_\mathrm{D}$ in $^{208}$Pb shows that present EDFs, relativistic as well as non-relativistic, cannot match the trend of $\alpha_\mathrm{D}$ from $^{208}$Pb to the Sn region.

The trends of $\alpha_\mathrm{D}$ along the Sn chain raise another intriguing question. 
The development with increasing nucleon number $A$ indicates a bend at $^{120}$Sn when deduced from the data. 
This is most likely a signature of shell effects implying that $\alpha_\mathrm{D}$ in open-shell nuclei is not only driven by bulk properties. 
Surprisingly, the EDF calculations with pairing produce a smooth, nearly linear increase mass number while calculations neglecting pairing qualitatively also show a pronounced kink.
The mismatch calls for a deeper analysis of the role of nuclear pairing. 

The present experimental results challenge the development of EDF parametrizations aiming for a systematic reproduction the dipole polarizability across the nuclear chart.  
Because of the strong correlation, such models will then provide improved predictions for the neutron skin thickness and parameters of the symmetry energy which, in turn, are important for extrapolation to star matter.
Combined with results expected from future studies of
neutron-rich unstable Sn isotopes using relativistic
Coulomb breakup with the R3B setup at FAIR \cite{aum20} which -- in contrast to the pioneering experiment by
Adrich et al.~\cite{adr05} -- will include information on the strength below neutron threshold, a unique set of data along an isotopic chain will be available to constrain isovector properties of nuclei and nuclear matter.

\section*{Acknowledgements}
The experiments were performed at RCNP under the experimental program E422. 
The authors thank the accelerator group for providing excellent beams.
We are indebted to J.~Piekarewicz for enlightening discussions on the impact of shell effects on the polarizability in the Sn isotope chain and to B.K.~Agrawal for providing calculations for the KDE functional.  
This work was funded by the Deutsche Forschungsgemeinschaft (DFG, German Research Foundation) -- Projektnummer 279384907 -- SFB 1245, by JSPS KAKENHI, Grant No.\ JP14740154 and by MEXT KAKENHI, Grant No.\ JP25105509.
G.C.\ and X.R.-M.\ acknowledge funding from the European Unions Horizon 2020 research and innovation program under grant agreement No.\ 654002.
C.A.B.\ was supported in part by U.S.\ DOE Grant No.\ DE-FG02-08ER41533 and U.S.\ NSF Grant No.\ 1415656.



\end{document}